\documentclass{jfm_arxiv}
\usepackage[utf8]{inputenc}
\usepackage[dvipsnames]{xcolor}
\usepackage{graphicx}
\usepackage{natbib}
\usepackage{hyperref}
\hypersetup{
    colorlinks = true,
    urlcolor   = blue,
    citecolor  = RoyalBlue,
}

\newcommand{\RomanNumeralCaps}[1]
\linenumbers

\newcommand{\D}{\ensuremath{\mathrm{D}}}
\newcommand{\A}{\ensuremath{\mathsf{A}}}

\newcommand{\vort}{\ensuremath{\omega}}

\renewcommand{\SS}{\ensuremath{\mathsf{S}}}
\renewcommand{\O}{\ensuremath{{\Omega}}}
\renewcommand{\ll}{\left\langle}
\newcommand{\rr}{\right\rangle}
\newcommand{\Or}[1]{\mathcal{O}\!\left(#1\right)}

\usepackage{mytools}

\colorlet{referee1}{Black}
    \colorlet{referee1b}{Black}
\colorlet{referee2}{Black}
\colorlet{referee3}{Black}

\title{A length scale for non-local multi-scale gradient interactions in isotropic turbulence}

\author{Miguel P. Encinar\aff{1}
\corresp{\email{miguel.pencinar@upm.es}}}

\affiliation{\aff{1}Department of Mechanical Engineering, Johns Hopkins University, Baltimore,
Maryland 21218, USA}

\date{}

\begin{document}
\maketitle

\begin{abstract}
    Three-dimensional turbulent flows enhance velocity gradients via strong non-linear interactions of the rate-of-strain tensor with the vorticity vector, and with itself. {\color{referee1b} For statistically homogeneous flows, their} total contributions to gradient production are related to each other by conservation of mass, and so are the total enstrophy and total dissipation. However, locally they do not obey this relation and have different (often extreme) values, and for this reason both production mechanisms have been subject to numerous studies, often decomposed in multiscale interactions. In general lines, their dynamics and contributions to the cascade processes and turbulent kinetic dissipation are different, which posses a difficulty for turbulence modelling. In this paper, we explore the consequence of the `Betchov' relations locally, and show that they implicitly define a length scale. This length scale is found to be about three times the size of the turbulent structures and their interactions. It is also found that while the non-locality of the dissipation and enstrophy at a given scale comes mostly from larger scales that do not cancel, the non-local production of strain and vorticity comes from multiscale interactions. An important consequence of this work is that isotropic cascade models need not distinguish between vortex stretching and strain self-amplification, but can instead consider both entities part of a more complex transfer mechanism, provided that their detailed point-value is not required and a local average of reasonable size is sufficient.
\end{abstract}

% \begin{keywords}
%     
% \end{keywords}
%
% {\bf MSC Codes }  {\it(Optional)} Please enter your MSC Codes here

\section{Introduction}
\label{sec:intro}
Non-locality is an essential feature of turbulent flows. The term `non-local' or
`non-local interactions' is used in turbulence research to refer to two distinct
phenomena, non-local interactions in physical space, and in Fourier space. This
paper focuses on the former. 

The ultimate source of non-locality is the incompressibility assumption, which
forces the velocity field to be solenoidal at each instant of time and point of space.
 Incompresibility acts through the pressure gradient term in the Navier--Stokes
 equation and thus it can be said that the pressure is responsible for non-locality.
The only way for the velocity field to be able to satisfy the incompressibility condition is to
communicate the movement of each fluid particle to all other particles instantly through the pressure.
For this reason it seems that the most straight-forward way of studying non-locality
is studying the characteristics of the pressure field. Indeed, some works explore the
characteristics of the isotropic pressure field directly \citep{bat:53,pum:94,tsu:ish:03}, although often they
are not concerned about non-locality in particular, although exceptions do exist \citep{wil:men:14, vla:wil:19}. 
%The reason is that many velocity statistics
% are not necessarily Galilean-invariant, and in isotropic turbulence, the velocity gradient field 
% is often the central object of study [TSINOBER]. The gradient field is traditionally decomposed
% in a rotational part or vorticity, and a stretching part or rate-of-strain [TENEKES]. 
Non-locality is instead typically studied in the context of enstrophy and dissipation amplification.
These quantities evolve through strain--vorticity interactions, which
are non-local, and through the pressure Hessian in the rate-of-strain evolution equation, which is also non-local. Both
terms have been extensively studied, particularly in the context of non-locality
\citep{ohk:kis:95, tsi:00, wil:men:14, els:17, men:11}. Moreover, the equations of the filtered velocity gradients can
be used to model inter-scale interactions, and non-locality here plays an important role, as inter-scale vortex stretching (VS) is often regarded as an important cascade indicator \citep{eyi:05}. %\citep{eyi:05,got:08}. 
 There are also indications that strain-self amplification (SSA) is equally (or up to three times more) important for the cascade rate \citep{tsi:00,car:bra:20,joh:20,yan:zho:xu:he:23} than VS.

Both VS and SSA are known to balance on average \citep{bet:56}, but locally can be very different from each other \citep{jim:wra:saf:rog:93,tsi:00}. The same can be said for the enstrophy and the non-dimensional dissipation, whose total magnitudes are related to each other. In principle, this allows one to study dissipation using the field of enstrophy as a proxy and, traditionally, research has favoured the study of the `simpler' vorticity vector over the more tedious rate-of-strain tensor.

The derivation of the evolution equations of the velocity gradients can, at most,
hide the effect of the pressure term in other kinds of non-local interactions, e.g. VS,
but cannot remove its impact. A particularly interesting attempt to separate local from non-local VS is \cite{ham:sch:dah:08}. This work exploits the integral transform from the vorticity to strain to produce local and non-local fractions of the VS. The same decomposition has been used by \cite{bua:pum:21} to analyse very high-Reynolds number turbulence, reaching the conclusion that most amplification comes from non-local interactions.
%I HAVE TO READ THEM. More recently WJ BOS uses a similar decomposition, but emmbeded in the dynamics. By separating both contributions to vortex stretching dynamically, they integrate modified N-S equations without non-local VS. The impact in the flow is ... I HAVE TO READ IT. 
These works have in common that a length scale needs to be chosen to separate local from non-local effects. Pressure is global in incompressible flows, making this length
scale nominally infinite. Thus, the concept of defining a length scale
associated to non-locality is ill-posed. However, even if the effects of pressure
need to be global, the magnitude of those effects across distance decays significantly, due to the properties of the Laplacian operator which appears in the equation for the pressure. This magnitude
is important because non-locality hinders our ability to understand gradient amplification and to produce sensible cascade models. 
This length scale is studied in \cite{vla:wil:19}, where the authors study the source term of the pressure fluctuations. Using correlations, they reach the conclusion that most of the non-locality arising from this term vanishes at distances larger than $20$ Kolmogorov units.

In the present paper, we take a similar approach to them, seeking to create a sensible definition of non-locality in terms of
velocity gradient interactions, and to measure the scale(s) at which non-locality is most intense, and more importantly, its scaling with the Reynolds number. Although `globality' is required to fully satisfy the continuity
equation, we argue that the main dynamical effects of continuity in the velocity gradients have a marked length scale, and
that said length scale is important for modelling flow dynamics. The main objective is to provide a sensible measure of the distances
in which non-local interactions are dominant, and when they may be neglected. We extend this analysis to the filtered velocity-gradient tensor, which allows us to generalise our conclusions from the dissipative range to the inertial one.

{\color{referee1}
The remainder of the paper is structured as follows. Section \S\ref{sec:numerics} introduces the databases used for this paper, and \S\ref{sec:method} recalls the equations that contain the quantities studied in this paper and their relation to Navier--Stokes. They are followed by section \S\ref{sec:manteca}, which shows all the results regarding gradient non-locality. Finally, \S\ref{sec:conclusions} closes with the final discussion and conclusions of the manuscript.
}

\section{Numerical databases}\label{sec:numerics}
We use databases of HIT obtained from direct numerical simulation (DNS) of the incompressible Navier--Stokes equations. Simulations use a standard fully dealiased spectral spatial discretisation coupled with a third-order semi-implicit Runge--Kutta time stepper. Turbulence is forced with constant energy input at the largest wavenumber corona, $|\boldsymbol{k}| < 2$, where $|\boldsymbol{k}|$ is the wavenumber magnitude. Table \ref{tab:kd} gathers the main characteristics of the simulations. {\color{referee2} The two lower $\Rey$ simulations use a smaller resolution, $k\eta_{\max{}} = 1.5$ than the \emph{de facto} standard of two. Although simulations as low as $k\eta_{\max{}} = 1$ have been used to gather useful statistics \citep[among others]{kan:isi:06}, high order moments, such as the statistics of the third invariant could be affected by this resolution. However, the good collapse of their statistics with the highest $\Rey$ simulation, which is computed at the standard resolution gives us confidence in the results of this paper.} More details about the numerical method and simulations are provided in \citet{car:vel:17}. 

\begin{table}
  \begin{center}
\def~{\hphantom{0}}
  \begin{tabular}{lcccccccc}
      Name & $N$  & $k\eta_{\max{}}$   & $L_E/\eta$ & $L_E/\lambda$ & $\epsilon L_E/u^{\prime3}$ & $Re_\lambda$ & $N_{\text{stat}}$& Symbol\\
      S256  & 256  & 1.5 & 119 & 5.03 & 0.524 & 144 & 640 & \includegraphics[height=2ex]{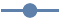}\\
      S512  & 512  & 1.5 & 213 & 7.12 & 0.462 & 231 & 160 & \includegraphics[height=2ex]{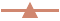}\\
      S1024 & 1024 & 2.0 & 333 & 9.80 & 0.494 & 298 & 40 & \includegraphics[height=2ex]{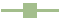}\\
  \end{tabular}
  \caption{Detail of the numerical simulations used in the paper. $N$ is the number of grid points in each of the periodic directions, $\eta$ is the Kolmogorov scale, $L_E$ the integral scale, $\lambda$ the Taylor microscale, $\epsilon$ is the total dissipation. All these quantities are computed as in \cite{bat:53}. $N_{\text{stat}}$ refers to the number of approximately independent snapshots used to compute the statistics in this paper.}
  \label{tab:kd}
  \end{center}
\end{table}

\section{Gradient amplification and non-locality}\label{sec:method}

% \subsection{Equations}
Consider the incompressible Navier--Stokes equations,
\begin{eqnarray}
    \partial_i u_i &=& 0\label{eq:cont}\\
    \D_t u_i \equiv \partial_t u_i + u_j \partial_j u_i &=&  -\partial_i p + \nu \partial_{kk} u_i,\label{eq:mom}
\end{eqnarray}
where $u_i$ is the velocity vector, $p$ the kinematic pressure and $\nu$ the
kinematic viscosity. $\D_t$ stands for the material derivative. The kinematic pressure
acts in \eqref{eq:mom} as a Lagrange multiplier, responsible for enforcing \eqref{eq:cont}
at each time and point in space. The only source of divergence in \eqref{eq:mom} is the term
$u_j\partial_ju_i$, which is the divergence of 
the Reynolds-stress tensor, $\partial_j\left(u_iu_j\right)$. A possible interpretation
% probably motivated by the fractional step procedure [FRACTIONAL STEP],
is that the Reynolds
stresses generate divergence, which is opposed and cancelled by the gradient of the pressure.
 However, this can be misleading, as both terms can be grouped together in the equation, and the non-linear term of Navier--Stokes equation is $\partial_j\left(u_iu_j + \delta_{ij}p\right)$, where $\delta$ is the Kronecker delta tensor. In this form, it can be seen explicitly
 that non-locality is a consequence of non-linearity. Taking the gradient of \eqref{eq:mom} yields,
\begin{equation}
    \D_t \A_{ij} = -\partial_{ij} p - \A_{ik}\A_{kj} + \nu\partial_{kk} \A_{ij}, \label{eq:Aij}
\end{equation}
where $\A_{ij} = \partial_j u_i$ is the velocity gradient tensor. The latter can be further
decomposed into a symmetric and skew-symmetric part, $\A_{ij} = \SS_{ij} + \O_{ij}$, where 
the symmetric $\SS_{ij}$ is the rate-of-strain tensor, and the skew-symmetric $\O_{ij}$ is the
rate-of-rotation tensor. The latter is related to the vorticity vector by $\vort_i = -1/2\varepsilon_{ijk}\O_{jk}$,
where $\boldsymbol{\vort}$ is the vorticity vector and $\varepsilon$ is the totally skew-symmetric tensor.
Evolution equations for $\SS$ and $\boldsymbol{\vort}$ can be constructed from \eqref{eq:Aij},
\begin{eqnarray}
    \D_t \SS_{ij} &=& -\partial_{ij} p + \SS_{ik}\SS_{kj} - \frac{1}{4}\left(\vort_i\omega_j - \boldsymbol{\omega}^2\delta_{ij}\right) + \nu\partial_{kk} \SS_{ij},\label{eq:Sij}\\
    \D_t \vort_i &=& \omega_j\SS_{ij} + \nu\partial_{kk} \omega_i,\label{eq:oi}
\end{eqnarray}
where $\boldsymbol{\vort}^2 = \omega_i\omega_i$. Non-locality appears in \eqref{eq:Sij} through the pressure Hessian, and in \eqref{eq:oi}
through the vortex-stretching term. The pressure is hidden in the vorticity equation through the strain-vorticity
interactions, as strain and vorticity are related to each other by a non-local integral transform \citep{ohk:94}. Note that both equations are non-local in the convective term too, as obtaining the velocity field from either the vorticity or the rate-of-strain involves using the 
incompressibility condition and inverting a second derivative. The trace of \eqref{eq:Sij} gives a recipe to compute
the pressure,
\begin{equation}
    \partial_{ii} p = \frac{1}{2}\boldsymbol{\vort}^2 - \SS^2,\label{eq:pres}
\end{equation}
where $\SS^2 = \SS_{ij}\SS_{ij}$. The pressure source term is proportional to the local imbalance of enstrophy and dissipation and, because for an isotropic flow the average pressure vanishes,
\begin{equation}
    \frac{1}{2}\left\langle{\vort^2}\right\rangle = \left\langle \SS^2\right\rangle,\label{eq:Qbalance}
\end{equation}
where the brackets are the spatial average. Note that for an isotropic flow with periodic or infinite boundaries the spatial average is sufficient to satisfy \eqref{eq:Qbalance}. Equation \eqref{eq:Qbalance} is an important non-local relation between the global enstrophy and the total dissipation. Equations for these two magnitudes can be obtained by contracting \eqref{eq:Sij} and \eqref{eq:oi} with $\SS$ and $\boldsymbol{\vort}$ respectively and taking the average,
\begin{eqnarray}
    \partial_t \ll\SS^2\rr &=&  - \ll \SS_{ij}\SS_{jk}\SS_{ki} \rr - \frac{1}{4}\ll\vort_i\SS_{ij}\omega_j\rr - \nu\ll\SS_{ij}\partial_{kk}\SS_{ij}\rr\label{eq:s2},\\
    \partial_t \ll\boldsymbol{\vort}^2\rr &=& \ll \omega_i\SS_{ij}\omega_j \rr - \nu\ll\omega_i\partial_{kk}\omega_i \rr\label{eq:o2}.
\end{eqnarray}
Production of enstrophy is achieved through the vortex stretching acting on the vorticity $\vort_i\SS_{ij}\omega_j$ (VS), which has to be positive on average in order to sustain turbulence. This term is a sink for the rate-of-strain magnitude, which in turn is produced by the term $\SS_{ij}\SS_{jk}\SS_{ki}$ or strain self amplification (SSA), which has to be negative on average. The expectancies of both quantities are linked by the celebrated `Betchov' relation \citep{bet:56},
\begin{equation}
    \ll \SS_{ij}\SS_{jk}\SS_{ki} \rr = \frac{3}{4}\ll\vort_i\SS_{ij}\omega_j\rr\label{eq:betchov}.
\end{equation}
% which is often invoked to recast the production of (filtered) gradients solely in
% terms of VS. 
This relation is purely kinematical, as only incompressibility, \textcolor{referee3}{isotropy} and the chain rule are required to prove it. \textcolor{referee1}{As stated in the introduction,
interpretations of \eqref{eq:betchov} are sometimes controversial. While some authors \citep[eg.][]{ohk:94, eyi:05} efectively use to recaset the production of (filtered) gradients solely in terms of VS, others \citep[eg.][]{tsi:01,car:bra:20,joh:20} question its local implications, as SSA and VS play, in principle, different roles in turbulence dynamics.}

Equations \eqref{eq:Qbalance} and \eqref{eq:betchov} are the two most important
kinematic relations which tie gradients and gradient production respectively across
the flow field. Moreover, they have been recently shown to be the only two homogeneous kinematic constrains \citep{car:wil:22}. They can be reworked exactly in terms of the average of the second and third invariants
of the velocity gradient tensor, $Q$ and $R$ \citep{cho:per:can:90},
\begin{eqnarray}
    \ll Q \rr &=& -\frac{1}{2}\ll \A_{ij}\A_{ji} \rr = \frac{1}{4}\ll \vort_i\omega_i - 2\SS_{ij}\SS_{ij} \rr = 0,\\
    \ll R \rr &=& -\frac{1}{3}\ll \A_{ij}\A_{jk}\A_{ki} \rr = -\frac{1}{3}\left(\SS_{ij}\SS_{jk}\SS_{ki} + \frac{3}{4}\vort_i\SS_{ij}\omega_j \right)=0.
\end{eqnarray}
Aside from the pressure Hessian, non-local kinematic constrains force the two invariants of the velocity gradient tensor to vanish. The first, which is the divergence of the velocity, is zero at each point in space due to incompressibility, and the spatial averages of the second or third are zero for an isotropic flow. In the following section we explore how cancellation of $Q$ and $R$ occurs in isotropic flows, but first we recall the typical way in which non-locality manifests in the velocity gradients.

\subsection{Evidence of non-locality}

\begin{figure}
    \raisebox{1em}{\includegraphics[width=0.30\textwidth]{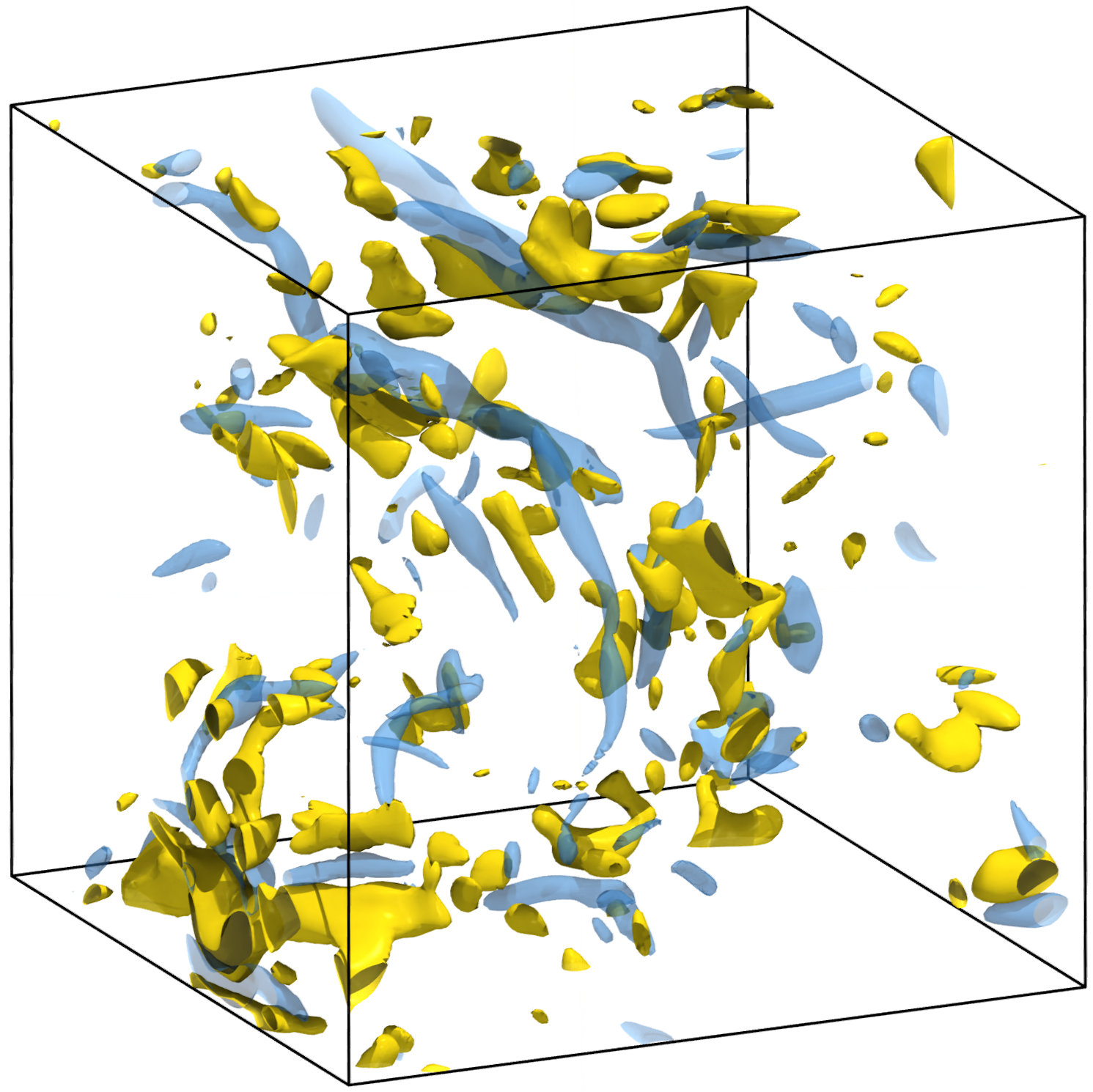}}\hfill%
    \includegraphics[width=0.32\textwidth]{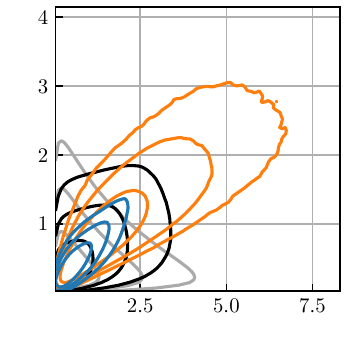}\hfill%
    \includegraphics[width=0.32\textwidth]{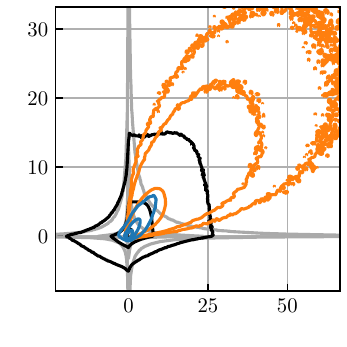}%
    \mylab{-7.0cm}{0cm}{$|{\boldsymbol \omega}|/{(\boldsymbol \omega^\prime)}^{1/2}$}%
    \mylab{-9.1cm}{2.15cm}{\rotatebox{90}{${|\mathsf{S}|}/{(\boldsymbol \omega^\prime)}^{1/2}$}}%
    \mylab{-2.35cm}{0cm}{${\boldsymbol \omega}\mathsf{S}{\boldsymbol \omega}/{(\boldsymbol \omega^\prime)}^{3/2}$}%
    \mylab{-4.5cm}{1.9cm}{\rotatebox{90}{$\mathsf{S}\mathsf{S}\mathsf{S}/{(\boldsymbol \omega^\prime)}^{3/2}$}}%
    \caption{(a) Subvolume of $(250\eta)^3$ of a snapshot of S256 (see table \ref{tab:kd}). Blue isosurfaces are  $|\boldsymbol{\omega}| > 4(\boldsymbol{\omega}^\prime)^{1/2} $ and yellow ones $|\mathsf{S}| > 4(\boldsymbol{\omega}^\prime)^{1/2} $. (b) Joint (p.d.f.)s of the enstrophy magnitude $|\boldsymbol{\omega}|$ and the dissipation magnitude $|\mathsf{S}|$. (c) Joint (p.d.f.)s of the VS  and the SSA. (b, c) Contours contain 90, 99 and 99.9\% of the probability mass. {\color{referee2} Black contours are the regular joint p.d.f. and grey contours are the product of the probabilities of the individual p.d.f. of the magnitudes. Ball coarse-grained quantities in a $12\eta$ radius are shown in blue and orange contours, the average is taken for the blue ones and the maximum is taken for the oranges.} {\color{referee1} (a,b,c) $\boldsymbol{\omega}^\prime$ stands for the variance of the vorticity.  }
    \label{fig:nonlocalrev}}
\end{figure}

Figure \ref{fig:nonlocalrev}(a) shows a 3D picture portraying non-locality in terms of the velocity gradients. It contains a small subvolume of isotropic turbulence, with contours for enstrophy and dissipation. The picture suggests that they are spatially related to each other, but it can be seen that they are not collocated.
These visual inspections are quantified by the joint probability density function (p.d.f.) of the enstrophy magnitude $|\boldsymbol{\omega}|$ and the dissipation magnitude $2|\mathsf{S}|$, \textcolor{referee1}{both normalised by the r.m.s. of the vorticity $(\boldsymbol{\omega}^\prime)^{1/2}$.} They are shown in black contours in figure \ref{fig:nonlocalrev}(b). Lighter grey contours contain {\color{referee1} the p.d.f. of both quantities if they were completely uncorrelated, obtained as the product of their probabilities, $P(\vort)P(\SS)$.} Visually, it reveals that a correlation of $C(\vort,\SS)\approx 0.57$ between the two fields exists. It also shows that strong events happen in different places, giving the p.d.f. its characteristic `square' shape.
% A better estimator of the information contained in the joint pdf is the Kullback--Leibler divergence [cite, $D_\mathrm{KL}$] of the pdf and its random counterpart, $R_\mathrm{KL}$,
% \begin{equation}
% R_\mathrm{KL}\left(|\boldsymbol{\omega}|, 2|\mathsf{S}|\right) = \int_\SSigma P(|\boldsymbol{\omega}|, 2|\mathsf{S}|)\log\left[\frac{P(|\boldsymbol{\omega}|,2|\mathsf{S}|)}{P(|\boldsymbol{\omega}|)P(2|\mathsf{S}|))}\right]\dd \sigma
% \end{equation}
% Intuitivelly, this quantity tells us how much information we obtain from knowledge of the joint PDF over the product of the individual ones. Its value is $R_\mathrm{KL}\left(|\boldsymbol{\omega}|, 2|\mathsf{S}|\right) \approx 0.173$, which is quite low. For example, a distribution of $2|\mathsf{S}|$  identical to $|\boldsymbol{\omega}|$ at each point of the flow would have a $R_\mathrm{KL}\approx 2.58$. Using either metric shows that strong enstrophy events are not collocated with strong strain events.
It has been known for some time that this is caused by strong vortices or `worms' surrounded
by strong rate-of-strain induced by them, and is often as an archetypical example of gradient non-locality \citep{jim:wra:saf:rog:93}.
%Their magnitudes are non-locally related, as show for example, by the expected value of the rate-of-strain given the vorticity and its converse, which are known to scale as power laws of each other [pumir].
Because vortices have a transversal length scale $\Or{\eta}$, the strain they induce must have the same length scale \citep{bur:48}. Thus, it is reasonable to expect that
both quantities should correlate at lengths of the same order. Blue contours show the joint pdf of $\langle|{\boldsymbol\omega}|^2\rangle_{12\eta}^{1/2}$ and $\langle 4|{\mathsf{S}}|^2\rangle_{12\eta}^{1/2}$. These quantities are the averaged squared norm of enstrophy and dissipation in spherical subvolumes of  radius $12\eta$. Both quantities correlate much better, with a correlation of $C\approx 0.86$%
%and a KL divergence of $R_\mathrm{KL} \approx 0.7$
. Finally, the orange contours show that this property does not only apply to events of average intensity but also to extreme ones. They represent the joint p.d.f. of the maximum value of the variables within the same {\color{referee1} spheres of radius $12\eta$}, and the correlation is similar to the one for the averages. {\color{referee2} This values are computed by taking the maximum value of each quantity instead of taking the average value within the ball. }As noted in \cite{bua:pum:22} there is a non-unity power-law relating the intensity of $|\boldsymbol{\omega}|$ and $|\mathsf{S}|$, which can be shown here as a different covariance between the blue and the orange contours.

Figure \ref{fig:nonlocalrev}(c) extends this analysis to the joint p.d.f. of VS and SSA, with its characteristic star-shape. Most of the conclusions from figure \ref{fig:nonlocalrev}(b) carry over here. Both VS and SSA can change sign and thus local averages can too. However, the probability of finding gradients being depleted diminishes significantly when the average is taken over volumes of $12\eta$, from 57\% and 50\% of the S512 domain having negative VS and positive SSA respectively, to 1.5\% and 1\% when $\langle\boldsymbol \omega \mathsf{S}\boldsymbol{\omega}\rangle_{12\eta}$ and $\langle\mathsf{S}\mathsf{S}\mathsf{S} \rangle_{12\eta}$ are considered. Correlations between the two variables improves by almost the same margins as $\langle|{\boldsymbol\omega}|^2\rangle_{12\eta}^{1/2}$ and $\langle 4|{\mathsf{S}}|^2\rangle_{12\eta}^{1/2}$. 

While figure \ref{fig:nonlocalrev} shows that non-locality has a strong mark in turbulent
velocity gradients, it also suggests that a sensible portion of the
non-local `Betchov' relations cancel over small distances within the dissipative range. Thus, cancellation implicitly defines a distance relating the components of the invariants, which we explore in \S\ref{sec:manteca}.

\section{Cancellation of the velocity-gradient invariants} % (fold)
\label{sec:manteca}
\begin{figure}
    \includegraphics[width=0.49\textwidth]{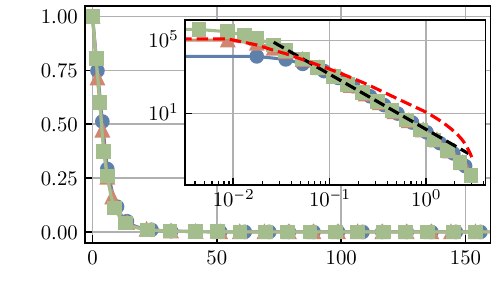}\hfill%
    \includegraphics[width=0.49\textwidth]{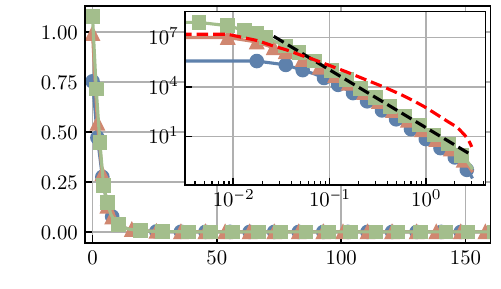}%
    \mylab{-1.9cm}{1cm}{\scalebox{0.7}{$r/L_E$}}%
    \mylab{-8.9cm}{1cm}{\scalebox{0.7}{$r/L_E$}}%
    \mylab{-3.25cm}{0cm}{$r/\eta$}%
    \mylab{-10.25cm}{0cm}{$r/\eta$}%
    \mylab{-7.0cm}{2.42cm}{$\frac{\ll R \rr_r^\prime}{(\ll Q \rr_0^\prime)^{3/2}}$}%
    \mylab{-13.75cm}{2.42cm}{$\frac{\ll Q \rr_r^\prime}{\ll Q \rr_0^\prime}$}%
    \mylab{-6.75cm}{3.5cm}{(b)}%
    \mylab{-13.5cm}{3.5cm}{(a)}%
    \caption{Variance of (a) $\ll Q \rr_r$ and (b) $\ll R \rr_r$ as a function of the integration radius for the three Reynolds numbers. The inset are a log-log plot of the same data, but normalised with the integral time scale \textcolor{referee2}{in the abscissa, and using the surrogate dissipation instead of $\ll Q \rr_0^\prime$ in the ordinate}. The red dashed line corresponds to the randomised test of S512. \label{fig:pdfqreta}}
\end{figure}
Figure \ref{fig:pdfqreta}(a) shows $\ll Q \rr_r^\prime$ for the three available Reynolds numbers, which stands for the variance $(\prime)$ of the coarse-grained $Q${\color{referee1} , defined as
\begin{equation}\ll Q \rr_r^\prime = \frac{1}{V}\int_V \ll Q \rr_r^2~\mathrm{d}x_i,
    \label{eq:Qstd}
\end{equation}
where $V$ stands for the computational domain. Note that the mean needs not to be substracted because $\ll Q \rr_r$ vanishes when integrated over the volume}.
The coarse-grain follows \cite{eyi:05}, averaging $Q$ over a sphere of radius $r$. When the radius goes to zero, we recover $Q^\prime$, and when $r\rightarrow \infty$, $\ll Q \rr_\infty^\prime$ must vanish due to incompressibility. Thus, $\ll Q \rr_r^\prime$ is a measure of how far the local enstrophy-dissipation balance is from the global one. Our results is very similar to the findings of \cite{vla:wil:19}, showing that most of the cancellation occurs within spheres of $r \approx 20\eta$, which encompass a vortex core and the strain it induces. The inset corresponds to the same curves, but in log-log scales. They are normalised with the surrogate dissipation \textcolor{referee1}{$D = (u^\prime)^{3/2}/L_E$}, which makes their tails collapse. The curves shown stop at $r/L_E \approx 1.6$ because we want to avoid the forced integral scale. The black line is a fitted power-law that corresponds to $p \approx r^{-3.02 \pm 0.1}$, and approximately holds for the three Reynolds numbers. The tails should be compared with the dashed red line, obtained from randomising the relative positions of the dissipation and enstrophy before computing $Q$. This artificial field of $Q$ still satisfies Betchov's equality, but assumes no spatial relation between its two components. Its power-law is $p \approx r^{-2.0}$ which corresponds to the growth of the surface of the integration sphere, and thus it cancels over larger length scales. Figure \ref{fig:pdfqreta}(b) is equivalent to \ref{fig:pdfqreta}(a) but shows $\ll R \rr_r$. It is normalised with $(Q^\prime)^{3/2}$, which results in intermittency effects showing at $r\rightarrow0$, even at our moderate Reynolds numbers. These effects cancel quickly with the coarse-graining, and $R$ roughly satisfies Betchov's equality for the same radius of $Q$. However, the fitted power-law is slightly different $p \approx r^{-3.5 \pm 0.15}$, which results in a faster cancellation. This figure shows that strong VS events are always accompanied by strong SSA events in their neighbourhood, and that the typical length scale of this interactions is of the same order of the interactions of dissipation and strain objects.

% \begin{itemize}
%     \item[v] explain the quantity a little 
%     \item[v] Wilzeck shows the same in a. B extends
%     \item[v] Q' squared as norm
%     \item[v] in R shows intermitence. It collapses very well
%     \item[v] Inset is log log to see the tail, its normalised with the surrogate dissipation.
%     \item[m] red is random, it shows the least possible restrictive version of betchov. The decay is much more "global"
%     \item[m] black is a power law. I Have to check how much in A and in B
% \end{itemize}

% We quantify
% the magnitude of this cancellation by computing the variance of the averages of the velocity
% gradient tensor over balls of a given radius, $\ll(\ll{Q}\rr_r)^2\rr$ and $\ll(\ll{R}\rr_r)^2\rr$. The former is
% a direct indicator of the cancellation of the pressure source term, while the later of the evolution of the velocity gradients. Figure \ref{fig:3lines} shows $\ll{Q}\rr_r(r)$ for the three Reynolds numbers available. It shows that 
% the three Reynolds numbers agree well among themselves, and that most of the cancellation takes place in distances smaller than $20\eta$. 
% These results are comparable to those of [Cite Wilzeck], which gives a similar cutoff using correlations. We extend these results to R.

% section manteca (end)
\subsection{Multiscale analysis of the Betchov invariants}\label{ssec:multi}

The results so far agree with those of \cite{vla:wil:19} and extend theirs to $R$, with similar conclusions. The cancellation of $Q$ is a consequence of vorticity generating strain or \emph{viceversa}, and can be interpreted as both entities being part of the same structure, i.e. a strong vortex with its self-induced strain \citep{els:mar:2010}. In the same way, the fact that the same cancellation holds for the VS and the SSA implies that the latter are paired somewhat locally, and that they probably come from the same dynamics, which generates them at the same time. In order for this to hold for the turbulent cascade, the same balance should hold at all inertial scales.

\begin{figure}
    \includegraphics[width=0.49\textwidth]{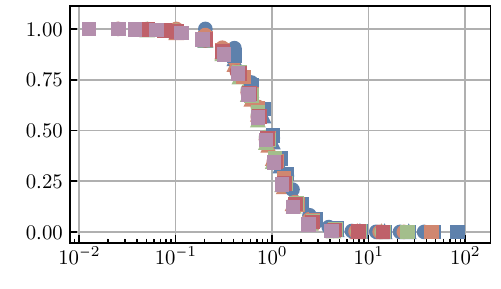}\hfill%
    \includegraphics[width=0.49\textwidth]{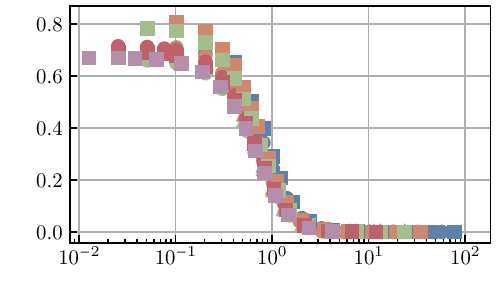}%
    \mylab{-3.25cm}{0cm}{$r/\rmDelta$}%
    \mylab{-10.25cm}{0cm}{$r/\rmDelta$}%
    \mylab{-7.0cm}{2.42cm}{$\frac{\ll \tilde R \rr_r^\prime}{(\ll Q \rr_0^\prime)^{3/2}}$}%
    \mylab{-13.75cm}{2.42cm}{$\frac{\ll \tilde Q \rr_r^\prime}{\ll Q \rr_0^\prime}$}%
    \mylab{-6.75cm}{3.5cm}{(b)}%
    \mylab{-13.5cm}{3.5cm}{(a)}%
    \caption{Variance of (a) $\ll \tilde Q \rr_r$ and (b) $\ll \tilde R \rr_r$ as a function of the integration radius for the three Reynolds numbers and various filters. See \S\ref{ssec:multi} for more details. \label{fig:pdfqretafil}}
\end{figure}

This is explored in figure \ref{fig:pdfqretafil}, which shows spherical averages similar to figure \ref{fig:pdfqreta}, but for the invariants of the filtered velocity fields. First, the velocity fields are filtered using a family of isotropic Gaussian kernels, with their widths separated in powers of two, and ranging from $10\eta$ to the integral scale,
\begin{equation}
    \tilde u_i(x_j) = {\color{referee1}\frac{1}{(\Delta\sqrt{\pi})^3}}\int_V \exp[-((x_j-\xi_j)/\Delta)^2]u_i(\xi_j)~\mathrm{d}\xi_j,
    \label{eq:ufilt}
\end{equation}
where $\Delta$ serves as the filter width.
The invariants of the filter velocity, $\tilde Q$ and $\tilde R$ represent the same incompressibility balances (and topology), but for the filtered velocity \citep{leu:swa:dav:12,loz:hol:jim:16,eyi:05,dan:men:18}. We repeat the coarse-graining of integrating both quantities in growing spheres of radius $r$, until we reach the integral scale. Although the fact that we are coarse-graining the fields twice may be striking, it should be clarified that both operations have different purposes. Filtering the velocity is an attempt to study the gradient interactions at the scale of the filter, which are otherwise hidden by the smaller scale gradients. Coarse-graining the invariants (which are non-linear functions of the velocity field) pursues the study of how local is the balance between the components that form the invariants. Figure \ref{fig:pdfqretafil}(a) shows $\ll \tilde Q \rr_r$ for every filter width that satisfies $30\eta < \rmDelta < 0.5L_E$, and it is considered to fall within the inertial range. These are five, six and seven filters for S256, S512 and S1024 respectively. They are normalised with $\tilde Q^\prime$. When the radius $r$ is normalised with $\rmDelta$, all the curves collapse very well. At $r/\rmDelta \approx 3$, more than 95\% of the cancellation has taken place. Similar numbers are given by figure \ref{fig:pdfqretafil}(b), which shows $\ll \tilde R \rr_r$. Again, it shows intermittentcy effects, with the ratio of $\tilde R^\prime$ over $(\tilde Q^\prime)^{3/2}$ being larger the farther the filter width is from the integral scale. Despite these effects, most of the cancellation still happens at distances smaller than three filter widths.

\begin{figure}
    \raisebox{0ex}{\includegraphics[width=0.49\textwidth]{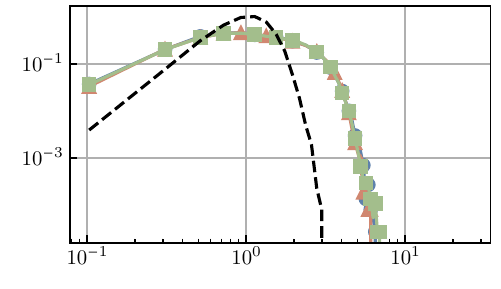}}\hfill%
    \raisebox{0ex}{\includegraphics[width=0.49\textwidth]{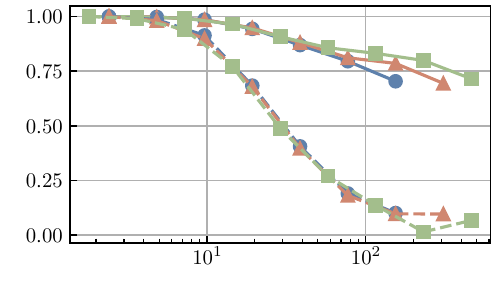}}%
    % \mngrid{6}{6}%
    \mylab{-10.25cm}{0cm}{$|\boldsymbol{d}|/(20\eta)$}%
    \mylab{-3.25cm}{0cm}{$\rmDelta/\eta$}%
    % \mylab{-5.5cm}{1.0cm}{${\boldsymbol \omega}$}%
    % \mylab{-4.5cm}{1.5cm}{${\mathsf{S}}$}%
    % \mylab{-1.75cm}{0.6cm}{${\boldsymbol \omega}$}%
    % \mylab{-2.0cm}{1.6cm}{${\mathsf{S}}$}%
    % \mylab{-3.75cm}{3.0cm}{${\mathsf{S}\mathsf{S}\mathsf{S}}$}%
    % \mylab{-3.4cm}{2.0cm}{${{\boldsymbol \omega}\mathsf{S{\boldsymbol \omega}}}$}%
    \mylab{-13.75cm}{2.42cm}{$p.d.f.$}%
    \mylab{-6.75cm}{2.42cm}{$C_{i}$}%
    \mylab{-6.75cm}{3.5cm}{(b)}%
    \mylab{-13.7cm}{3.5cm}{(a)}%
    \mylab{-3.0cm}{1.5cm}{$C_R$}%
    \mylab{-2.0cm}{3.4cm}{$C_Q$}%
    \caption{(a) P.d.f. of the distance between the minima of $\langle Q \rangle|_{\Delta/\eta=30}$ and $\langle R \rangle|_{\Delta/\eta=30}$.
    %THE EXPECTANCY IS 1.63; FOR QQ or RR is 0.5.
    (b) Correlations between the filtered and coarse-grained velocity-gradient invariances. Lines with circles are for the second invariant and dashed lines with squares are for the third.}
    \label{fig:excess}
\end{figure}

{\color{referee2} The fact that the cancellation or `Betchov' length scale is the same for both invariants may mislead the reader to think that they occur within the same spheres, i.e. that they are part of the same `structure'. However, we found that}
the cancellation of $\ll Q \rr$ and $\ll R \rr$ is not collocated, meaning that the fact that enstrophy and dissipation are in local equilibrium does not imply that VS and SSA are, and \emph{viceversa}.  Figure \ref{fig:excess} explores their relation by showing the p.d.f. of the minimum distance across local minima of $\langle Q \rangle_{r/\eta=20}$ and $\langle R \rangle_{r/\eta=20}$. They are computed as follows. First, the local minima are identified by looking for zeroes of the gradient of the $\langle Q \rangle_{r/\eta=20}^2$ and $\langle R \rangle_{r/\eta=20}^2$ fields (Note that the mean of these fields are the variances as defined in \eqref{eq:Qstd}). Then, for each of the two fields, we retain the minima which are closest to zero in magnitude (closer to local equilibrium), until the density of retained minima in the field is a constant. The latter is such that the average distance between minima, or to the minima of the other quantity is $20\eta$, which is the radius of the coarse-graining. Changing this density to twice or half its value does not change the results, as long as they are renormalised with the new density. {\color{referee2} These sets of points represent the centres of regions where either $Q$ or $R$ are in local equilibrium}. The minimum distance between these minima gives us information about the organisation of the regions where the Betchov relation is satisfied locally \textcolor{referee2}{for each quantity}. The random Poisson model {\color{referee1} for the distance between randomised points} is shown as a dashed line, which would match the data if no relation existed between the set of local minima of $\langle Q \rangle_{r/\eta=20}$ and the local minima of $\langle R \rangle_{r/\eta=20}$. The data however does not fit a Poisson process, and instead shows a much larger tail to the right, falling from a  flat plateau in the range of one to three times the coarse-graining length. The expectancy of the distance is 1.63$~r$, and it is consistent for the three Reynolds numbers. This organisation shows that the equilibrium of both quantities is related to each other, and that they tend to be further apart than a random process. {\color{referee2} Although physically unsound, the reader may find useful to think of both sets of points as if they were repelling or avoiding each other.} If we interpret the flow in terms of the evolution equations for the gradients, then dissipation and enstrophy represent structures or eddies, and VS and SSA their interactions. The fact that they are not collocated highlights that interactions of multiple `$Q$' objects are require to produce `$R$' ones.
% \begin{itemize}
%     \item explain the filtered Q and R and then integrated
%     \item it is the same that unfiltered but with delta
%     \item in R again shows the pile up of the cascade. It still collapses in a length scale
%     \item This suggsts that at a given scale, there are Q objects and R objects.
%     \item Not shown is Q-R' (maybe if it fits). Is not zero. This shows that Q and R are not collocated even when avereaged. Suggests that different Q objects interact with each other producing R objects are interacting.
% \end{itemize}

Consider now figure \ref{fig:excess}(a). It shows the correlation between the filtered invariants and the coarse-grained ones,
\begin{equation}
    C_Q(\rmDelta) = \dfrac{\int_V \tilde Q (x_i) \ll Q\rr_\rmDelta (x_i) ~\mathrm{d}x_i}{
    \left(\int_V \ll Q\rr_\rmDelta (x_i) \ll Q\rr_\rmDelta (x_i) ~\mathrm{d}x_i\int_V \tilde Q (x_i) \tilde Q (x_i) ~\mathrm{d}x_i
\right)^{1/2}}, 
    \label{eq:correlationq}
\end{equation}
and equivalently for $R$. The filter width of the Gaussian is chosen to match $\rmDelta$, so both quantities have comparable scales. These two quantities, $\tilde Q$ and $\ll Q \rr_\Delta$, although they may seem similar, have very different meanings. While $\tilde Q(x_i; \rmDelta)$ represents the local balance of dissipation and enstrophy for scales larger than $\Delta$ at the position $x_i$, $\ll Q\rr(x_i; \rmDelta)$ represents the {excess of $Q$ needed to be cancelled in order to satisfy the kinematical relation locally, after scales below $\rmDelta$ have been cancelled}.  
\textcolor{referee1}{This is not an effect of the difference between the coarse-graining and the filtering kernel, but one that comes from the step at which this operation is performed. While $\ll Q\rr_\Delta(x_i; \rmDelta) = -1/2
\ll \A_{ij}\A_{ji}\rr_\Delta$ is constructed from the total velocity field, $\tilde Q = -1/2\tilde\A_{ij}\tilde\A_{ji}$ comes from the contraction of the filtered velocity gradient tensor}. The correlations are shown to collapse as a function of $\rmDelta/\eta$, except for the scales $\rmDelta/L_E \sim 1$ which do so when normalised with the integral scale. These large scales show approximately a constant correlation coefficient for all three Reynolds numbers and approximately equal to $0.75$. For $Q$, the correlation is approximately unity for scales smaller than $10\eta$, and decays smoothly into the large scales. The fact that this correlation coefficient is large implies that most of the structure of $\ll Q \rr$ comes from $\tilde Q$. This can be interpreted as the enstrophy-dissipation balance at scales larger than the filter being most responsible for the lack of local kinematic balance. It suggest a picture, when paired with figure \ref{fig:pdfqretafil}(a), in which at each scale, local balance is approximately satisfied within a few filter widths, and the excess comes from the scales above the filter. %This  model is compatible with the power-law observed in the tail of figure \ref{fig:pdfqreta}(a), and with multiplicative cascades. 

A similar analysis can be done for $\tilde R$ and $\ll R \rr$, and it is shown using dashed lines. The correlation decays far from the dissipative range, and the fields of both quantities are essentially uncorrelated at the integral scale. This result shows that the self-scale interactions at larger scales cannot be the main contribution to the balance of gradient amplification and that multiscale contributions, i.e. amplification of gradients at one scale due to the gradients at a different one, are the most important contribution, in agreement with \cite{bua:pum:21}. Multi-scale interactions are part of the turbulent cascade, which can be expressed in terms of the filtered gradient production terms \citep{joh:20}. Although evaluating those terms are out of the scope of the present paper, they can be compared with the excess of Betchov's balance using the multiscale procedure of \cite{yan:zho:xu:he:23}, for example.\vspace{-5pt}

\section{Discussion and Conclusions}\label{sec:conclusions}
We have shown that most of the kinematical balance between the enstrophy and the dissipation, and between the VS and the SSA is satisfied within spheres of radius of the order of $20\eta$. For their counterparts computed from the filter velocities, cancellation is achieved at lengths of the order of three filter widths, for as long as the filter width lies within the inertial range. The remaining imbalance correlates well with the filtered source in the case of the second invariant but notably worse for the third. 

{\color{referee2} If the filter-width is considered representative of the size of the structures captured by the filter, then the `Betchov' length scale is about three times the size of an `eddy', both for the second and the third invariants. The former can be interpreted as representative of turbulent structures comprised of vorticity surrounded by strain, balancing in three times the typical width of either the vorticity or the strain structures. The latter represents the interaction of vorticity and strain structures, resulting in the production of both VS and SSA, which also balance within three times their size. The same arguments hold for the unfiltered invariants, as $7$--$10\eta$ is the typical size of vortices and strain sheets.}

The conclusion is that for isotropic flows, or isotropic cascade models, the assumption that dissipation can be replaced by enstrophy, or strain-self amplification by vortex stretching (e.g \citep{eyi:05}) is physically reasonable, even if dynamically they are completely different quantities. This statement only holds if the point value of the components is not considered dynamically relevant and only its effect in a small neighbourhood is considered. The present results suggest that VS and SSA should be modelled or studied as part of a larger dynamical mechanism `the vortex-strain amplification' which, while necessarily not local, it is approximately balanced in a small neighbourhood of a reasonable size $r/\rmDelta \approx 3$. This idea is also supported by the fact that their proportion can be attributed solely to kinematical relations \citep{yan:zho:xu:he:23}. A sensible model compatible with the successful ideas of the multiplicative cascades is that enstrophy-dissipation objects, i.e. vortices and their induced strain, interact with each other; generating VS-SSA mechanisms at various scales. This is supported by two findings. First, the fact that the local balance of $Q$ is not collocated with the local balance of $R$, and that their distribution is not independent from each other. It is also compatible with the power-laws of figure \ref{fig:pdfqreta}.
Second, the present results suggest that most of the excess imbalance of enstrophy and dissipation comes from larger scales. However, that is not true for the VS-SSA pair, which implies that the imbalance must come from interscale interactions. Additional work should clarify whether, for example, these interactions happen in a neighbourhood in scale space \citep{car:vel:17}, or are fully global in this sense.

Finally, we would like to note that while it is perfectly possible to measure the contribution of both VS and SSA to the turbulent cascade, it seems of less dynamical relevance than their combined contributions acting over a broader region in isotropic flows. This distinction, however, may be necessary to model non-isotropic flows under strong shear, where the balance may not be satisfied at the length scales shown here. 

\backsection[Acknowledgments]{The author would like to thank Dr. Alberto Vela-Mart\'in  and Beatriz J\'imenez-Carrasco for their useful comments in a preliminary version of this manuscript, and the three anonymous referees  who helped polish this paper to its final version. I would also like to thank Prof. J. Jim\'enez for providing the computational resources used for this work.}
\backsection[Funding]{The author received no specific grant from any funding agency. The computational resources have been provided by the ERC.2020.AdG.101018287, ``CausT''}
\backsection[Declaration of interest]{The author reports no conflict of interest.}
\bibliographystyle{jfm_arxiv}

\bibliography{jfm}

%% End of file `jfm2esam.bib'.

\end{document}